\documentclass[conference
]{IEEEtran}

\usepackage{color}
\usepackage{amssymb}
\usepackage{graphicx}
\usepackage{multirow}
\usepackage{algorithmic}
\usepackage{epstopdf}

\usepackage[scriptsize]{subfigure}
\graphicspath{{.}}  

\newtheorem{theorem}{Theorem}

\newcommand{\datvectrv}{\mathbf{X}}
\newcommand{\outvectrv}{\mathbf{Y}}

\newcommand{\entropy}{\mathit{H}}
\newcommand{\mutualinfo}{\mathit{I}}

\newcommand{\test}{T}
\newcommand{\universe}{n}
\newcommand{\defective}{d}

\newcommand{\threshold}{\Delta} 
\newcommand{\error}{\epsilon}
\newcommand{\logerror}{\delta}

\begin{document}

\title{Non-adaptive probabilistic group testing with noisy measurements: Near-optimal bounds with efficient algorithms}

\author{\IEEEauthorblockN{Chun Lam Chan, Pak Hou Che and Sidharth Jaggi}
\IEEEauthorblockA{Department of Information Engineering\\
The Chinese University of Hong Kong}
\and
\IEEEauthorblockN{Venkatesh Saligrama}
\IEEEauthorblockA{Department of Electrical and Computer Engineering\\
Boston University}}

\maketitle

\begin{abstract}
We consider the problem of detecting a small subset of defective items from a large set via {\it non-adaptive ``random pooling'' group tests}. We consider both the case when the measurements are noiseless, and the case when the measurements are noisy (the outcome of each group test may be independently faulty with probability $q$). Order-optimal results for these scenarios are known in the literature. We give information-theoretic lower bounds on the query complexity of these problems, and provide corresponding computationally efficient algorithms that match the lower bounds up to a constant factor. To the best of our knowledge this work is the first to explicitly estimate such a constant that characterizes the gap between the upper and lower bounds for these problems.
\end{abstract}

\IEEEpeerreviewmaketitle

\section{Introduction}


The goal of {\it group testing} is to identify a small unknown subset ${\mathcal D}$ of defective items embedded in a much larger set ${\mathcal N}$ (usually in the setting where $|{\mathcal D}|$ is much smaller than $|{\mathcal N}|$, {\it i.e.}, $|{\mathcal D}|$ is $o(|{\mathcal N}|)$). This problem was first considered by Dorfman \cite{Dorfman1943} in scenarios where multiple items in a group can be simultaneously tested, with a binary output depending on whether or not a ``defective'' item is presented in the group being tested. In general, the goal of group testing algorithms is to identify the defective set with as few measurements as possible. As demonstrated in \cite{Dorfman1943} and future work, with judicious grouping and testing, far fewer than the trivial upper bound of $|{\mathcal N}|$ may be required to identify the set of defective items. 


In this work our model has four important assumptions. 
\begin{itemize}
\item {\it Non-adaptive group testing}: The set of items being tested in each test is required to be independent of the outcome of every other test. This restriction is often useful in practice, since this enables parallelization of the testing process. It also allows for an automated testing process (whereas the procedures and especially the hardware required for {\it adaptive} group testing may be significantly more complex). Furthermore, it is known (for instance~\cite{CGT}, \cite{saligrama09}) that adaptive group testing algorithms do not improve upon non-adaptive group-testing algorithms by more than a constant factor in the number of tests required to identify the set of defective items. 

\item {\it ``Small-error" group testing:} Our algorithms are allowed to have a ``small'' probability of error. It is known (for instance~\cite{saligrama09}) that zero-error algorithms require significantly more tests asymptotically (in the number of defective items) than algorithms that allow asymptotically small errors. 

\item {\it ``Noisy" measurements:}
In addition to the {\it noiseless} group-testing problem specified by the above, we also consider the ``noisy'' variant of the problem, wherein the result of each test may differ from the true result (in an independent and identically distributed manner) with a certain pre-specified probability $q$. \footnote{An alternative model involving ``worst-case" errors has also been considered in the literature (for instance~\cite{Macula1997217}), wherein the designed group-testing algorithm is required to be resilient to {\it all} noise patterns wherein at most a fraction $q$ of the results differ from their true values, rather than the probabilistic guarantee we give against {\it most} fraction-$q$ errors. This is analogous to the difference between combinatorial coding-theoretic error-correcting codes (for instance Gilbert-Varshamov codes~\cite{GV}) and probabilistic information-theoretic codes (for instance~\cite{Shannon:48}). In this work we concern ourselves only with the latter, though it is possible that our techniques can also be used to analyzed the former.} Since the measurements are noisy, the problem of estimating the set of defective items is more challenging, and is known to require more tests. \footnote{{\bf We wish to highlight the difference between {\it noise} and {\it errors}. We use the former term to refer to noise in the outcomes of the group-test, regardless of the group-testing algorithm used. The latter term is used to refer to the error due to the estimation process of the group-testing algorithm.}}

\item {\it Computationally efficient and near-optimal algorithms:} Most algorithms in the literature focus on optimizing the number of measurements required -- in some cases, this leads to algorithms that may not be computationally efficient to implement (for {\it e.g.}~\cite{saligrama09}).  In our work we require our algorithms be both computationally efficient and near-optimal in the number of measurements required. 
\end{itemize}

In the literature, order-optimal upper and lower bounds on the number of tests required are known for the problems we consider (for instance~\cite{saligrama09}, \cite{Sejdinovic2010}). In both the noiseless and noisy variants, the number of measurements required to identify the set of defective items is known to be $T = \Theta(d\log(n))$ -- here $n = |{\mathcal N}|$ is the total number of items and $d = |{\mathcal D}|$ is the size of the defective subset. However, in the noisy variant, the number of tests required is in general a constant factor larger than in the noiseless case (where this constant $\beta$ is independent of both $n$ and $d$, but may depend on the noise parameter $q$ and the allowable {\it error-probability} $\delta$ of the algorithm.

However, to the best of our knowledge, prior to this work no explicit characterization has been given of the actual number of measurements required (rather than just order-optimal results). In particular, we analyze two algorithms that we call Combinatorial Basis Pursuit (CBP), and Combinatorial Orthogonal Matching Pursuit (COMP),\footnote{This choice of nomenclature is motivated by two popular Compressive Sensing decoding algorithms, respectively Basis Pursuit, and Orthogonal Matching Pursuit -- as we note in Section~\ref{subsec:prior_work}, the decoding algorithms we analyze in this work might be viewed as combinatorial analogues of those well-analyzed algorithms.} that have both been previously considered in the group-testing literature (under different names) for both noiseless and noisy scenarios (see, for instance, \cite{NGT}). We provide explicit upper bounds on $\beta(q,\delta)$ for both these algorithms. Further, we also provide corresponding lower bounds on $\beta(q,\delta)$ for {\it any} group-testing algorithms. These upper and lower bounds are asymptotically independent of both $n$ and $d$. The lower bounds are information-theoretic, and the upper bounds are derived from a detailed analysis of CBP and COMP under both the noiseless and noisy scenarios. In general, the bounds resulting from the analysis of the algorithms match our simulations to a high degree, which indicates that the bounds we derive are not too slack.

This paper is organized as follows. In Section \ref{sec:Background}, we introduce the model and corresponding notation, and describe the algorithms analyzed in this work. In Section \ref{sec:Result}, we describe the main results of this work. Sections~\ref{sec:lwrbnd} and \ref{sec:uprbnd} contain the analysis respectively our information-theoretic lower bounds, and of the group-testing algorithms considered. Our simulation results are presented in Section \ref{sec:Simulation}.

\subsection{Prior Work}
\label{subsec:prior_work}

Dorfman~\cite{Dorfman1943} first considered the group-testing problem during World-War II with regards to testing soldiers for syphilis. Since then, a large body of literature has considered the problem (see for instance the book by~\cite{CGT}). 

In this work we focus on non-adaptive algorithms. Here we can further subdivide algorithms according to whether errors (that decay asymptotically to zero with large $n$) are allowed or not in the reconstruction algorithm, and, orthogonally, whether the measurements are noisy are not.

If errors are not allowed in the group-testing algorithm, it is known that at least $\Omega(d^2 \log(n))$ tests are required in both noiseless and noisy scenarios (which may be considerably larger than the $\Theta(d\log(n))$ bounds that are known (for instance \cite{saligrama09} and \cite{Sejdinovic2010}) for the ``small-error" scenario. Further, in the noisy scenario, if no errors are allowed in the reconstruction algorithm, only noise patterns with an absolute bound on the total number of noisy measurements can be handled.\footnote{This is because in the ``usual" noise model, wherein each measurement may be noisy with a certain probability, there is a non-zero probability that an arbitrary fraction of the measurements are corrupted in an arbitrarily bad manner. In this case no group-testing algorithm can hope to decode with zero-error.} For these reasons, we choose to focus on algorithms in which a small probability of error is allowed -- the reader interested in zero-error algorithms is encouraged to look at \cite{CGT}, \cite{DyaRyk82}, \cite{Dyachkov1989}, \cite{Cheng2009:1862-4472:297}, \cite{Chen20091581}.

The works closest to ours are those of \cite{saligrama09} and \cite{Sejdinovic2010}. The former analyzes the performance of certain group-testing algorithms in both noiseless and noisy settings information-theoretically, and proves order-optimality. However, only order-optimal (rather than explicit) bounds on the number of tests required are provided, and also the algorithms analyzed are not provided, and also the algorithms analyzed are not computationally efficient. The work of \cite{Sejdinovic2010} proposes a belief-propogation decoding rule to improve the computational efficiency of the algorithms of \cite{saligrama09}, but no proof of correctness is provided. In contrast, in this work we provide the first explicit bounds on computationally efficient group-testing algorithms.

Information-theoretic lower bounds on the number of tests required are folklore -- some instances of these bounds for some models are provided in~\cite{Dyachkov1989}. Since we were unable to find a specific reference covering all cases for our model, we also prove our lower bounds in Section~\ref{sec:lwrbnd}. 

There are intriguing connections between the two algorithms we consider, and corresponding Compressive Sensing (CS) algorithms. In particular, Basis Pursuit has been well-analyzed in the CS literature (for instance~\cite{CandesTao, Donoho}), as has Orthogonal Matching Pursuit (for instance~\cite{Tropp07signalrecovery}). The primary difference between those algorithms and the ones considered here is that in CS all measurements are over the real field ${\mathbb R}$, whereas in group-testing the measurements are modeled instead as an OR of AND clauses (hence the term ``Combinatorial").

\section{Background}
\label{sec:Background}

\subsection{Model and Notation}

\begin{figure}[htbp]
	\centering
		\includegraphics[width=60mm]{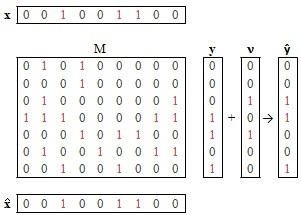}
		\caption {An example demonstrating a typical non-adaptive group-testing setup. The $T \times n$ binary group-testing matrix represents the items being tested in each test, the length-$n$ binary input vector ${\mathbf x}$ is a weight $d$ vector encoding the locations of the $d$ defective items in ${\cal D}$, the length-$T$ binary vector ${\mathbf y}$ noiseless result denotes the outcomes of the group tests in the absence of noise, the length-$T$ binary noisy result vector ${\mathbf {\hat{y}}}$ denotes the actually observed noisy outcomes of the group tests, as the result of the noiseless result vector being perturbed by the length-$T$ binary noise vector ${\mathbf \nu}$. The length-$n$ binary estimate vector $\hat{\mathbf x}$ represents the estimated locations of the defective items.}
		\label{fig:gt}
\end{figure}

A set ${\cal N}$ contains $n$ items, of which an unknown subset ${\cal D}$ are said to be ``defective".\footnote{In this work, as is common (see for example \cite{springerlink:10.1007/BF01609876}), we assume that the number $d$ of defective items in ${\cal D}$, or at least a good upper bound on them, is known {\it a priori}. If not, other work (for example \cite{SOBEL01041975}) considers non-adaptive algorithms with low query complexity that help estimate $d$.} The goal of group-testing is to correctly identify the set of defective items via a minimal number of ``group tests", as defined below (see Figure~\ref{fig:gt} for a graphical representation).

Each row of a $T \times n$ binary {\it group-testing matrix} $M$ corresponds to a distinct test, and each column corresponds to a distinct item. Hence the items that comprise the group being tested in the $i$th test are exactly those corresponding to columns containing a $1$ in the $i$th location. The method of generating such a matrix $M$ is part of the design of the group test -- this and the other part, that of estimating the set ${\cal D}$, is described in Section~\ref{subsec:algorithm}.

The length-$n$ binary {\it input} vector ${\mathbf x}$ represents the set ${\cal N}$, and contains $1$s exactly in the locations corresponding to the items of ${\cal D}$. The locations with ones/defective items are said to be {\it positive} -- the other locations are said to be {\it negative}. We use these terms interchangeably.

The outcomes of the {\it noiseless} tests correspond to the length-$T$ binary {\it noiseless result} vector ${\mathbf y}$, with a $1$ in the $i$ location if and only if the $i$th test contains at least one defective item. 

The observed vector of test outcomes in the {\it noisy} scenario is denoted by the length-$T$ binary {\it noisy result} vector ${\mathbf {\hat{y}}}$ -- the probability that each entry of ${\mathbf y}$ differs from the corresponding entry in ${\mathbf {\hat{y}}}$ is $q$, where $q$ is the {\it noise parameter}. The locations where the noiseless and the noisy result vectors differ is denoted by the length-$T$ binary {\it noise vector} ${\mathbf \nu}$, with $1$s in the locations where they differ. 

The estimate of the locations of the defective items is encoded in the length-$n$ binary {\it estimate vector}, with $1$s in the locations where the group-testing algorithms described in Section~\ref{subsec:algorithm} estimate the defective items to be.


The {\it probability of error} of any group-testing algorithm is defined as the probability (over the input vector ${\mathbf x}$, group-testing matrix $M$, and noise vector ${\mathbf{\nu}}$) that the estimated vector differs from the input vector.

\subsection{Algorithms}
\label{subsec:algorithm}

We now describe the CBP and COMP algorithms in both the noiseless and noisy settings. The algorithms are specified by the choices of encoding matrices and decoding algorithms.

\noindent {\bf 1. NOISELESS ALGORITHMS}
\hspace*{\fill} \\

\noindent {\bf Combinatorial Basis Pursuit (CBP)}:

The $T \times n$ group-testing matrix $M$ is defined as follows. A {\it group sampling parameter} $g$ is chosen (the exact values of $T$ and $g$ are code-design parameters to be specified later). Then, the $i$th row of $M$ is specified by sampling with replacement from the set $[1,\ldots,n]$ exactly $g$ times, and setting the $(i,j)$ location to be one if $j$ is sampled at least once during this process, and zero otherwise.\footnote{Note that this process of sampling each item in each test with replacement results in a slightly different distribution than if the group-size of each test was fixed {\it a priori} and hence the sampling was ``without replacement" in each test. (For instance, in the process we define, each test may, with some probability, test fewer than $g$ items.) The ``without replacement" process is a perhaps more natural way of defining tests, and also experimentally seems to result in slightly better performing algorithms. However, the corresponding analysis is significantly trickier, and we have been unable to find closed form expressions for such ``without replacement" sampling. The primary advantage of analyzing the ``with replacement" sampling is that in the resulting group-testing matrix every entry is then chosen {\it i.i.d.}.}

The decoding algorithm proceeds by using {\it only} the tests which have a negative (zero) outcome, to identify all the non-defective items, and declaring all other items to be defective. If $M$ is chosen to have enough rows (tests), each non-defective test should, with significant probability, appear in at least one negative test, and hence will be appropriately accounted for. Errors (false positives) occur when at least one non-defective item is not tested, or only occurs in positive tests ({\it i.e.,} every test it occurs in has at least one defective item). The analysis of this type of algorithm comprises of estimating the trade-off between the number of tests and the probability of error.

More formally, for all tests $i$ whose measurement outcome $y_i$ is a zero, let ${\mathbf m_i}$ denote the corresponding $i$th row of $M$, and ${\mathbf {m(y)}}$ denote the length-$n$ binary vector which has $1$s in exactly those locations where there is a $1$ in at least one ${\mathbf m_i}$. The decoder sets ${\mathbf {\hat{x}}}$ as ${\mathbf {1-m(y)}}$, {\it i.e.,} it has zeroes where ${\mathbf {m(y)}}$ has ones, and vice versa.

The rough correspondence between this algorithm and Basis Pursuit (\cite{CandesTao, Donoho}) arises from the fact that, as in Basis Pursuit, the decoder attempts to find a ``sparse" solution ${\mathbf {\hat{x}}}$ that can generate the observed vector ${\mathbf y}$.

\begin{figure}[htbp]
	\centering
		\includegraphics[width=30mm]{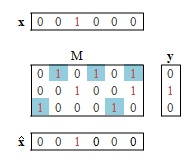}
		\caption {An example demonstrating the CBP algorithm. Based on only on the outcome of the negative tests (those with output zero), the decoder estimates the set of non-defective items, and ``guesses" that the remaining items are defective.}
		\label{fig:cbp}
\end{figure}

\noindent {\bf Combinatorial Orthogonal Matching Pursuit (COMP)}:

The $T \times n$ group-testing matrix $M$ is defined as follows. A {\it group selection parameter} $p$ is chosen (the exact values of $T$ and $p$ are code-design parameters to be specified later). Then, i.i.d. for each $(i,j)$, the $(i,j)$th element of $M$ is set to be one with probability $p$, and zero otherwise.

The decoding algorithm columns-wise, instead of row-wise as in CBP. It attempts to match the columns of $M$ with the result vector ${\mathbf y}$. That is, if a particular column $j$ of $M$ has the property that all locations $i$ where it has ones {\it also} corresponds to ones in $y_i$ in the result vector, then the $j$th item ($x_j$) is declared to be defective (positive). All other items are declared to be non-defective (negative). 

Note that this decoding algorithm never has false negatives, only false positives. A false positive occurs when {\it all} locations with ones in the $j$th column of $M$ (corresponding to a non-defective item $j$) are ``hidden" by the ones of other columns corresponding to defectives items. That is, let columns $j$ and some other columns $j_1,\ldots,j_k$ of matrix $M$ be such that for each $i$ such that $m_{i,j} = 1$, there exists an index $j'$ in $\{j_1,\ldots,j_k\}$ for which $m_{i,j'}$ also equals $1$. Then if each of the $\{j_1,\ldots,j_k\}$th items are defective, then the $j$th item will also always be declared as defective by the COMP decoder, regardless of whether or not it actually is. The probability of this event happening becomes smaller as the number of tests $T$ become larger.
Hence, as in CBP, the analysis of this type of algorithm comprises of estimating the trade-off between the number of tests and the probability of error.

The rough correspondence between this algorithm and Orthogonal Matching Pursuit (\cite{Tropp07signalrecovery}) arises from the fact that, as in Orthogonal Matching Pursuit, the decoder attempts to match the columns of the group-testing matrix with the result vector.

\begin{figure}[htbp]
	\centering
		\includegraphics[width=60mm]{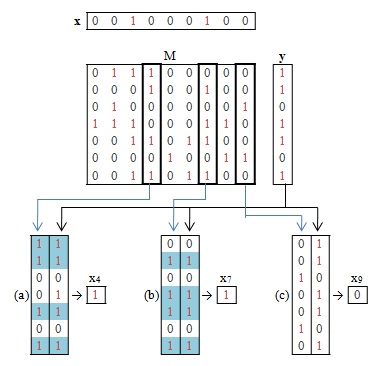}
		\caption {An example demonstrating the COMP algorithm. The algorithm matches columns of $M$ to the result vector. As in (b) in the figure, since the result vector ``contains" the $7$th column, then the decoder declares that item to be defective. Conversely, as in (c), since there is no such containment of the last column, then the decoder declares that item to be non-defective. However, sometimes, as in (a), an item that is truly negative, is ``hidden" by some other columns corresponding to defective items, leading to a false positive. }
		\label{fig:comp}
\end{figure}




\noindent {\bf 2. NOISY ALGORITHMS}
\hspace*{\fill} \\

\noindent {\bf Noisy Combinatorial Basis Pursuit (NCBP)}:

Let $K$ be design parameters to be specified later. To generate the $M$ for the NCBP algorithm case, each row of $M$ from the noiseless CBP algorithm is repeated $K$ times. The decoder declares the result of each a particular set of $K$ successive tests to be positive if at least $K/2$ of the tests in that group are positive, and else declares each such test to actually be negative. The decoder then uses the noiseless CBP algorithm to estimate ${\cal D}$.


\noindent {\bf Noisy Combinatorial Orthogonal Matching Pursuit (NCOMP)}

Finally, we consider the algorithm whose analysis is the major result of this work. In the noisy COMP case, we relax the sharp-threshold requirement in the original COMP algorithm that the set of locations of ones in any column of $M$ corresponding to a positive item be {\it entirely} contained in the set of locations of ones in the result vector. Instead, we allow for a certain number of ``mismatches" -- this number of mismatches depends on both the number of ones in each column, and also the noise parameter $q$.

Let $p$ and ${\Delta}$ be design parameters to be specified later. To generate the $M$ for the NCOMP algorithm case, each element of $M$ is selected i.i.d. with probability $p$ to be $1$. 

The decoder proceeds as follows, For each column $i$, we define the {\it indicator set} ${\cal T}_i$ as the set of indices $j$ in that column where $m_{i,j} = 1$. We also define the {\it matching set} ${\cal S}_i$ as the set of indices $j$ where both $\hat{y}_j = 1$ (corresponding to the noisy result vector) and $m_{i,j}=1$.

Then the decoder uses the following ``relaxed" thresholding rule. If $|{\cal S}_i| \geq |{\cal T}_i|(1-q(1+\Delta))$, then the decoder declares the $i$th item to be defective, else it declares it to be non-defective.


\begin{figure}[htbp]
	\centering
		\includegraphics[width=60mm]{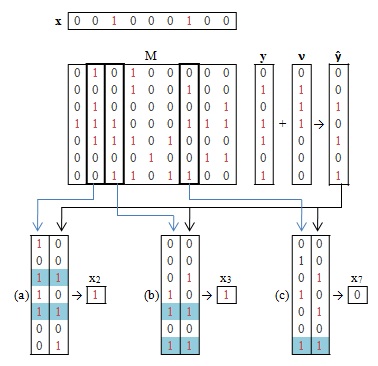}
		\caption {An example demonstrating the NCOMP algorithm. The algorithm matches columns of $M$ to the result vector {\it up to a certain number of mismatches} governed by a threshold. In this example, the threshold is set so that the number mismatches be less than the number of matches. For instance, in (b) above, the $1$s in the third column of the matrix match the $1$s in the result vector in two locations (the $5$th and $7$th rows), but do not match only in one location in the $4$th row (locations wherein there are $0$s in the matrix columns but $1$s in the result vector do not count as mismatches). Hence the decoder declares that item to be defective, which is the correct decision. \protect \\ However, consider the false negative generated for the item in (c). This corresponds to the $7$th item. The noise in the $2$nd, $3$rd and $4$th rows of ${\mathbf \nu}$ means that there is only one match (in the $7$th row) and two mismatches ($2$nd and $4$th rows) -- hence the decoder declares that item to be non-defective. \protect \\ Also, sometimes, as in (a), an item that is truly negative, has a sufficient number of measurement errors that the number of mismatches is reduced to be below the threshold, leading to a false positive.}
		\label{fig:ncomp}
\end{figure}

\section{Main Results}
\label{sec:Result}

\subsection{Lower Bounds}

We first provide information-theoretic lower bounds on the number of tests required by {\it any} group-testing algorithm. While we believe these bounds to be ``common knowledge" in the field, we have been unable to pinpoint a reference that gives an explicit lower bound on the number of tests in terms of the acceptable probability of error of the group-testing algorithm. For the sake of completeness, so we can benchmark our analysis of the algorithms we present later, we state and prove the lower bounds here. All logarithms in this work are assumed to be binary.

\begin{theorem}
\label{thm1} \label{thm:1}
[Folklore] Any group-testing algorithm with noiseless measurements that has a probability of error of at most $\error$ requires at least $(1-\error)\defective \log(\universe / \defective)$ tests.
\end{theorem}

In fact, the corresponding lower bounds can be extended to the scenario with noisy measurements as well.


\begin{theorem}
\label{thm2} \label{thm:2}
[Folklore] Any group-testing algorithm that has measurements that are noisy i.i.d. with probability $q$ and that has a probability of error of at most $\error$ requires at least $[(1-\error)d\log(n/d)]/(1-H(q))$ tests.\footnote{Here $H(.)$ denotes the binary entropy function.}
\end{theorem}


{\bf Note:} Our assumption that $d = o(n)$ implies that the bounds in Theorem~\ref{thm1} and~\ref{thm2} are $\Omega{\left ( d\log(n) \right )}$.

\subsection{Upper Bounds}
The main contributions of this work are explicit computations of the number of tests required to give a desired probability of error via computationally efficient algorithms. In both the noiseless and noisy case, we consider two types of algorithms (CBP and COMP). Both these algorithms have been considered before in the literature (for instance, see \cite{NGT}), but to the best of our knowledge ours is the first work to explicitly compute the tradeoff between the number of tests required to give a desired probability of error, rather than giving order of magnitude estimates of the number of tests required for a ``reasonable" probability of success.

\begin{theorem}
\label{thm3} \label{thm:3}
CBP with error probability at most $n^{-\logerror}$ requires no more than $2 (1+\logerror) ed\ln n$ tests.
\end{theorem}

\begin{theorem}
\label{thm4} \label{thm:4}
COMP with error probability at most $n^{-\logerror}$ requires no more than $ed(1+\logerror)\ln(n)$ tests.
\end{theorem}

Note that Theorems~\ref{thm3} and~\ref{thm4} is commensurate with the corresponding lower bound in Theorem~\ref{thm2}.

Translating these algorithms into the noisy measurement case is non-trivial. One approach is to repeat the tests in CBP, leading to NCBP and the following theorem.

\begin{theorem}
\label{thm5} \label{thm:5}
NCBP with probability of error at most $n^{-\logerror}$ requires no more than $2 e (1+\logerror) 2(\ln \ln n + \ln d + \logerror \ln n + 1 + \ln (2(1+\logerror)))(1-2q)^{-2}d\log(n)$
tests.
\end{theorem}

Note that this is asymptotically worse than the corresponding lower bound in Theorem~\ref{thm2}. We therefore provide the main result of this paper, 

\begin{theorem}
\label{thm6} \label{thm:6}
NCOMP with error probability at most $n^{-\logerror}$ requires no more than $4.36(\sqrt{\logerror}+\sqrt{1+\logerror})^2(1-2q)^{-2}d\log n$ tests.
\end{theorem}

Note that the corresponding upper bound differs from the lower bound by a factor that is at most $4.36(\sqrt{\logerror}+\sqrt{1+\logerror})^2(1-2q)^{-2}$, which is a function only of $q$ and $\logerror$. for ``small'' $q$ this quantity is ``small''.

\section{Proof of lower bounds}\label{sec:lwrbnd}


We begin by noting that $\datvectrv \rightarrow \outvectrv \rightarrow \hat{\outvectrv} \rightarrow \hat{\datvectrv}$ ({\it i.e.} the input vector, noiseless result vector, noisy result vector, and the estimate vector) forms a Markov chain. By standard information-theoretic definitions we have
\begin{eqnarray}
	\entropy(\datvectrv) &=& \entropy(\datvectrv | \hat{\datvectrv}) + \mutualinfo(\datvectrv ; \hat{\datvectrv}) \nonumber
\end{eqnarray}
Since $\datvectrv$ is uniformly distributed over all length-$\universe$ and $\defective$-sparse data vectors, $\entropy (\datvectrv) = \log |\mathcal{X}| = \log {\universe \choose \defective}$. By Fano's inequality, $\entropy(\datvectrv | \hat{\datvectrv}) \leq 1 + \epsilon \log {\universe \choose \defective}$. Also, we have $\mutualinfo(\datvectrv ; \hat{\datvectrv}) \leq \mutualinfo(\outvectrv ; \hat{\outvectrv})$ by the data-processing inequality. 
Finally, note that $$I({\hat{\mathbf{Y}}};{\hat{\mathbf{Y}}})  \leq\sum_{i=1}^T \left [ H({\hat{Y}_i}) - H({\hat{Y}_i}|Y_i) \right ]$$
since the first term is maximized when each of the ${\hat{Y}_i}$ are independent, and because the measurement noise is memoryless. For the BSC($q$) noise we consider in this work, this summation is at most $T(1-H(q))$ by standard arguments.\footnote{This technique also holds for more general types of discrete memoryless noise -- for ease of presentation, in this work we focus on the simple case of the Binary Symmetric Channel.}

Combining the above inequalities, we obtain
\begin{eqnarray}
	(1 - \epsilon) \log {\universe \choose \defective} & \leq & 1 + \test (1 - \entropy{(q)}) \nonumber
\end{eqnarray}
Also, by standard arguments via Stirling's approximation~\cite{CoverandThomas}, $\log{{n}\choose {d}}$ is at least $d\log(n/d)$. Substituting this gives us the desired result
\begin{eqnarray}
	\test	& \geq & \frac{1 - \epsilon}{1 - \entropy{(q)}} \log {\universe \choose \defective} \nonumber \\
			& \geq & \frac{1 - \epsilon}{1 - \entropy{(q)}} \defective \log \left( \frac{\universe}{\defective} \right). \nonumber
\end{eqnarray}
\hfill $\Box$

\section{Proofs of upper bounds} \label{sec:uprbnd}
\label{sec:Analysis}

\subsection{Noiseless Group Testing}

\vspace{0.1in}
\noindent {\bf Proof of Theorem~\ref{thm:3}}:

The Coupon Collector's Problem (CCP) is a classical problem that considers the following scenario. Suppose there are $n$ types of coupons, each of which is equiprobable. A collector selects coupons (with replacement) until he has a coupon of each type. What is the distribution on his stopping time? It is well-known (\cite{wiki:ccp}) that the expected stopping time is $n\ln n + \Theta(n)$. Also, reasonable bounds on the tail of the distribution are also known -- for instance, it is known that the probability that the stopping time is more than $\chi n \ln n$ is at most $n^{-\chi + 1}$.


Analogously to the above, we view the group-testing procedure of CBP as a Coupon Collector Problem. Consider the following thought experiment. Suppose we consider any test as a length-$g$ {\it test-vector } \footnote{Note that this test-vector is different from the binary length-$n$ vectors that specify tests in the group testing-matrix, though there is indeed a natural bijection between them.} whose entries index the items being tested in that test (in this view, repeated entries are allowed in this vector). Due to the design of our group-testing procedure in
CBP, the probability that any item occurs in any location of such a vector is uniform and independent. In fact this property (uniformity and independence of the value of each entry of each test) also holds {\it across} tests. Hence, the items in any subsequence of $k$ tests may be viewed as the outcome of a process of selecting a single chain of $gk$ coupons. This is still true even if we restrict ourselves solely to the tests that have a negative outcome. The goal of CBP may now be viewed as the task of collecting {\it all} the {\it negative} items. This can be summarized in the following equation

\begin{equation}
\label{eq:1}
Tg\left ( \frac{n-d}{n} \right )^g \geq (n-d)\ln (n-d) .
\end{equation}

Modifying (\ref{eq:1}) to obtain the corresponding tail bound on $T$ takes a bit more work. The right-hand side is then modified to $\chi (n-d) \ln (n-d)$ (which corresponds to the probability that all types of coupons have not been collected if these many total coupons have been collected is at most $(n-d)^{-\chi+1}$). The left-hand side is multiplied with $(1-\rho)$, where $\rho$ is a design parameter to be specified by Chernoff's bound on the probability that the actual number of items in the negative tests is smaller than $(1-\rho)$ times the expected number. By Chernoff's bound this is at most $\exp \left (-\rho^2T \left ( \frac{n-d}{n}  \right )^g\right )$. Taking the union bound over these two low-probability events gives us that the probability that
\begin{equation}
\label{eq:2}
(1-\rho)Tg\left ( \frac{n-d}{n} \right )^g \geq \chi (n-d)\ln (n-d)
\end{equation}
does {\it not} hold is at most
\begin{equation}
\exp \left (-\rho^2 T \left(\frac{n-d}{n}\right)^g \right ) + (n-d)^{-\chi + 1}.
\label{motherfucker}
\end{equation}


So, optimizing for $g$ in (\ref{eq:1}) and substituting $g^{\ast} = 1/\ln \left ( \frac{n}{n-d} \right )$ into (\ref{eq:2}), and noting that $\left ( \frac{n-d}{n}\right )^{g^{\ast}}$ equals $e^{-1}$, we have


\begin{eqnarray}
	T & \geq & \frac{\chi}{1 - \rho} \frac{(n-d) \ln (n-d)}{g^{*} \left(\frac{n-d}{n}\right)^{g^{*}}} \nonumber \\
	&=& \frac{\chi}{1 - \rho} \frac{(n-d) \ln (n-d)}{\frac{1}{\ln \left(\frac{n}{n-d}\right)} e^{-1}} \nonumber \\
	&=& \frac{\chi}{1 - \rho} \frac{(n-d) \ln (n-d) \ln \left(\frac{n}{n-d}\right)}{ e^{-1}}. \label{bb}
\end{eqnarray}

Using the inequality $\ln(1+x) \geq x - x^2/2$ with $x$ as $d/(n-d)$ simplifies the RHS of (\ref{bb}) to
\begin{equation}
	T \geq \frac{\chi}{1 - \rho} e \left( d - \frac{d^2}{2(n-d)} \right) \ln (n-d).
\label{bb2}
\end{equation}

Choosing $T$ to be greater than the bound in (\ref{bb2}) can only reduce the probability of error, hence
choosing
\begin{equation}
T \geq \frac{\chi}{1-\rho}e d\ln(n-d)
\label{iloveyou}
\end{equation}
\noindent still implies a probability of error at most as large as in (\ref{motherfucker}).

Choosing $\rho = \frac{1}{2}$ and substituting (\ref{iloveyou}) into (\ref{motherfucker}) implies, for large enough $d$, the probability of error $P_e$ satisfies
\begin{eqnarray}
		 P_e &\leq & e^{-\frac{\delta^2 \chi}{1 - \delta}d \ln (n-d)} + (n-d)^{-\chi+1} \nonumber \\
		 &=& (n-d)^{-\frac{\delta^2}{1-\delta} \chi d} + (n-d)^{-\chi+1} \nonumber \\
		 &\leq & 2(n-d)^{-\chi+1}. 
\end{eqnarray}



Taking $2(n-d)^{-\chi+1} = n^{-\delta}$, we have $\chi = \delta \frac{\log n}{\log (n-d)} + \frac{1}{\log (n-d)} + 1$. For large $n$, $\chi $ approaches $ \delta + 1$.

Therefore, the probability of error is at most $n^{-\delta}$, with sufficiently large $n$, the following number of tests suffice to fulfil the terms of the theorem
\[
	T \geq 2 (1+\delta) e d \ln n.
\]
\hfill $\Box$
\vspace{0.1in}

\noindent {\bf Proof of Theorem~\ref{thm4}:}

As noted in the discussion on COMP in Section~\ref{subsec:algorithm}, the error-events for the algorithm correspond to false positives, when a column of $M$ corresponding to a non-defective item is ''hidden" by other columns corresponding to defective items. To calculate this probability, recall that each entry of $M$ equals one with probability $p$, i.i.d. Let $j$ index a column of $M$ corresponding to a non-defective item, and let $j_1,\ldots, j_d$ index the columns of $M$ corresponding to defective items. Then the probability that $m_{i,j}$ equals one, and at least one of $m_{i,j_1}, \ldots, m_{i,j_d}$ {\it also} equals one is $p(1-(1-p)^d)$. Hence the probability that the $j$th column is hidden by a column corresponding to a defective item is $\left ( 1-p(1-p)^d \right )^T$. Taking the union bound over all $n-d$ non-defective items gives us that the probability of false positives is bounded from above by 
\begin{equation}
\label{eq:COMP_false_positive}
P_e = P_e^+ \leq (n-d)\left ( 1-p(1-p)^d \right )^T.
\end{equation}
By differentiation, optimizing (\ref{eq:COMP_false_positive}) with respect to $p$ suggests choosing $p$ as $1/d$. Substituting this value back into (\ref{eq:COMP_false_positive}), and setting $T$ as $\beta d \ln n$ gives us
\begin{eqnarray}
P_e & \leq & (n-d)\left ( 1-\frac{1}{de} \right )^{\beta d \ln n} \nonumber \\
		& \leq & (n-d)e^{-\beta e^{-1}\ln n} \nonumber \\
		& \leq & n^{1 - \beta e^{-1}}.
\end{eqnarray}
Choosing $\beta = (1+\delta)e$ thus ensures the required decay in the probability of error. Hence choosing $T$ to be at least $(1+\delta)ed\ln n$ suffices to prove the theorem. \hfill $\Box$.

\vspace{0.1in}

\subsection{Noisy Group Testing}

We now consider the harder problem of group testing when the measurements are noisy. First, just as a benchmark, we consider using the noiseless CBP algorithm with each test repeated identically $K$ times, where $K$ is a parameter to be determined so as to ensure a probability of error that can be made to decay asymptotically in $n$.

\vspace{0.1in}
\noindent {\bf Proof of Theorem~\ref{thm5}:}

Since each test has a probability $q$ of giving the wrong result, by the Chernoff bound the probability that more than the threshold number of tests give the incorrect result is at most $e^{-2K(\frac{1}{2}-q)^2}$. Hence by the union bound, repeating each of the $T$ tests $K$ times, the probability that the decoder makes an error is at most 
\begin{equation}
\label{eq:ncbp1}
P_e\leq T\left (e^{-2K(\frac{1}{2}-q)^2}  \right ).
\end{equation}

Substituting $T$ as $\beta d \log n$ implies that for the probability (\ref{eq:ncbp1}) to approach zero asymptotically in $n, K$ must be at least 
\begin{eqnarray}
K & \geq & \frac{2(\logerror \ln n + \ln T)}{(1-2q)^2}
\nonumber \\
& \geq & \frac{2(\ln \ln n + \ln d + \logerror \ln n + 1 + \ln (2(1+\logerror)))}{(1-2q)^2}.
\end{eqnarray}.
\hfill $\Box$

\noindent {\bf Note:} As noted earlier, the number of tests required by NCBP is larger than the corresponding lower bound in Theorem~\ref{thm:2} by a factor that is larger than any constant.

\vspace{0.1in}

\noindent {\bf Proof of Theorem~\ref{thm6}:}


Due to the presence of noise, both false positives and false negatives may occur in the noisy COMP algorithm -- the overall probability of error is the sum of the probability of false positives and that of false negatives. We set $p=\alpha / d$ (where $\alpha$ is a code-design parameter to be determined later) and $T= \beta d \log n$. We first calculate the probability of false positives by computing the probability that more than the expected number of ones get flipped to zero in the result vector in locations corresponding to ones in the column indexing the defective item. This can be computed as 
\begin{eqnarray}
{P}_e^- & = & \bigcup_{i=1}^{d} P\left ( |{\cal T}_i|=t \right )P\left ( |{\cal S}_i|< |{\cal T}_i|(1-q(1+\threshold )) \right )
\nonumber \\
& \stackrel{}{\leq} & d\sum_{t=0}^{T}{T \choose t}p^t(1-p)^{T-t}\label{justification:na}
\\ && \: \sum_{r=t-t(1-q(1+\threshold ))}^{t}{t \choose r}q^r(1-q)^{t-r}
\\
& \stackrel{\label{justification:nb}}{\leq} & d\sum_{t=0}^{T}{T \choose t}p^t(1-p)^{T-t}e^{-2t(q\threshold)^2}
\\
& \stackrel{\label{justification:nc}}{=} & d\left ( 1-p+pe^{-2(q\threshold)^2} \right )^T
\\
& \stackrel{\label{justification:nd}}{=} & d\left ( 1-\frac{\alpha}{d}+\frac{\alpha}{d}e^{-2(q\threshold)^2} \right )^{\beta d \log n}
\\
& \stackrel{\label{justification:ne}}{\leq} & de^{-\alpha \beta \left ( 1-e^{-2(q\threshold)^2 } \right )\log n}
\\
& \stackrel{}{\leq} & de^{-\alpha \beta (1-e^{-2})(q\threshold)^2 \log n}\label{justification:nf}
\end{eqnarray}
Here, as in Section~\ref{subsec:algorithm}, ${\cal T}_i$ denotes the locations of ones in the $i$th column of $M$. Inequality (\ref{justification:na}) follows from the union bound over the possible errors for each of the defective items, with the first summation accounting for the different possible sizes of ${\cal T}_i$, and the second summation accounting for the error events corresponding to the number of one-to-zero flips exceeding the threshold chosen by the algorithm. Inequality (\ref{justification:nb}) follows from the Chernoff bound. Equality (\ref{justification:nc}) comes from the binomial theorem. Equality (\ref{justification:nd}) comes from substituting in the values of $p$ and $T$. Inequality (\ref{justification:ne}) follows from the leading terms of the Taylor series of the exponential function. Inequality (\ref{justification:nf}) follows from an appropriate linear lower bound to the concave function $1-e^{-x}$.


For the requirement that the probability of false negatives be at most $n^{-\logerror}$ to be satisfied implies that $\beta^{-}$ (the bound on $\beta$ due to this restriction) be at least $(\alpha (1-e^{-2})(q\threshold)^2)^{-1}((\ln d / \ln n)+ \logerror)\ln 2$. Since $d=o(n)$ this converges to 
\begin{equation}
\label{beta_n}
\beta ^-> \frac{\logerror \ln 2}{\alpha (1-e^{-2})(q\threshold)^2}.
\end{equation}


We now focus on the probability of false positives. In the noiseless CBP algorithm, the only way a false positive could occur was if all the ones in a column are hidden by ones in columns corresponding to defective items. In the noisy CBP algorithm this still happens, but in addition noise could also lead to a similar masking effect. That is, even in the $1$ locations of a non-defective column not hidden by other defective columns, measurement noise flips enough zeroes to ones so that the decoding threshold is exceeded, and the decoder declares that particular item to be defective. See Figure~\ref{fig:ncomp}(a) for an example.

Hence we define a new quantity $a$, which denotes the probability for any $(i,j)$th location in $M$ that a $1$ in that location is ``hidden by other columns {\it or} by noise". It equals 
\begin{eqnarray}
a=1-[(1-q)(1-p)^d+q(1-(1-p)^d)].
\nonumber 
\end{eqnarray}
To facilitate our analysis, as $\left ( 1+\frac{x}{n} \right )^n\leq e^x$ for $n>0$, we bound 
\begin{IEEEeqnarray}{rCl}
a & = & 1-q-(1-p)^d(1-2q)
\nonumber \\
& =  & 1-q-(1-\frac{\alpha }{d})^d(1-2q)
\nonumber \\
& \geq  & (1-q) - e^{-\alpha}(1-2q).
\label{false positive}
\end{IEEEeqnarray}
The probability of false positives is then computed in a similar manner to that of false negatives as in (\ref{justification:na})--(\ref{justification:nf}).
\begin{IEEEeqnarray}{rCl}
P_e^+ & = & \bigcup_{i=1}^{n-d} P\left ( |{\cal T}_i|=t \right )P\left ( |{\cal S}_i|\geq |{\cal T}_i|(1-q(1+\threshold )) \right )
\nonumber \\
& \leq & (n-d)\sum_{t=0}^{T}{T \choose t}p^t(1-p)^{T-t} \nonumber \\ && \: \sum_{r=t(1-q(1+\threshold ))}^{t}{t \choose r}a^r(1-a)^{t-r}
\nonumber \\
& \leq  & (n-d)\left ( 1-p+pe^{-2((1-q(1+\threshold ))-a)^2} \right )^T
\label{justification:pa} \\
& \leq  & (n-d) \nonumber \\ && \: \left ( 1-p+pe^{-2[e^{-\alpha}(1-2q) - \threshold q)]^2} \right )^T
\label{justification:pb} \\
& \leq & (n-d) \nonumber \\ && \: e^{-\alpha \beta \left ( 1-e^{-2[e^{-\alpha}(1-2q)-\threshold q]^2} \right )\log n}
\nonumber \\
& \leq & (n-d) \nonumber \\ && \: e^{-\alpha \beta (1-e^{-2})(e^{-\alpha}(1-2q)-\threshold q)^2\log n}
\end{IEEEeqnarray}
Note that for the Chernoff bound to applicable in (\ref{justification:pa}), $1-q(1+\threshold)>q$. Equation (\ref{justification:pb}) follows from substituting the bound derived on $a$ in (\ref{false positive}) into (\ref{justification:pa}).
For the requirement that the probability of false positives be at most $n^{-\logerror}$ to be satisfied implies that $\beta^{+}$ (the bound on $\beta$ due to this restriction) be at least $(\alpha (1-e^{-2})(e^{-\alpha}(1-2q)-\threshold q)^2)^{-1}((\ln (n-d))/(\ln n)+ \logerror)\ln 2$. Since $d=o(n)$ this converges to
\begin{equation}
\label{beta_p}
\beta ^+> \frac{\left ( 1 + \logerror \right ) \ln 2}{\alpha (1-e^{-2})(e^{-\alpha}(1-2q)-\threshold q)^2}.
\end{equation}


Note that $\beta $ must be at least as large as $\max\{\beta^{-},\beta^{+}\}$ so that both (\ref{beta_n}) and (\ref{beta_p}) are satisfied.

When the threshold in the noisy COMP algorithm is high ({\it i.e.,} $\threshold$ is small) then the probability of false negatives increases; conversely, the threshold being low ($\threshold$ being large) increases the probability of false positives. Algebraically, this expresses as the condition that $\threshold > 0$ (else the probability of false negatives is significant), and conversely to the condition that $1-q(1+\threshold) > a$ (so that the Chernoff bound can be used in (\ref{justification:pa})) -- combined with (\ref{false positive}) this implies that $\threshold \leq e^{-\alpha}(1-2q)/q$. For fixed $\alpha$, each of (\ref{beta_n}) and (\ref{beta_p}) as a function of $\threshold$ is a reciprocal of a parabola, with a pole the corresponding extremal value of $\threshold$. Furthermore, $\beta^{-}$ is strictly increasing and $\beta^{+}$ is strictly decreasing in the region of valid $\threshold$ in $(0,e^{-\alpha}(1-2q)/q)$. Hence the corresponding curves on the right-hand sides of (\ref{beta_n}) and (\ref{beta_p}) intersect within the region of valid $\threshold$, 
and a good choice for $\beta$ is at the $\threshold$ where these two curves intersect. Let 
\begin{equation}
\gamma = (\ln d + \logerror \ln n)/(\ln (n-d) + \logerror \ln n).
\label{logerror}
\end{equation}
(Note that for large $n$, since $d = o(n)$, $\gamma$ approaches $\logerror/(1+\logerror)$.) Then equating the RHS of (\ref{beta_n}) and (\ref{beta_p}) implies that the optimal $\threshold^\ast$ satisfies
\begin{IEEEeqnarray}{rCl}
\frac{\ln 2}{\alpha (1-e^{-2})(e^{-\alpha}(1-2q)-\threshold q)^2} & = & \frac{\gamma \ln 2}{\alpha (1-e^{-2})\threshold^2 q^2}
\label{eq:betas}
\end{IEEEeqnarray}
Simplifying (\ref{eq:betas}) gives us that 
\begin{equation}
\label{epsilon_star}
\threshold^* = \frac{e^{-\alpha}(1-2q)}{q(1+\gamma^{-1/2})}.
\end{equation}
Substituting (\ref{epsilon_star}) into (\ref{beta_p}) we see that the resulting function can be viewed as $e^{2\alpha}/\alpha$ times factors that are independent of $\alpha$.
Optimizing this with respect to $\alpha$ indicates that the minimal value of $\beta$ occurs when $\alpha = 0.5$.

Substituting these values of $\alpha$, $\gamma$ and $\threshold$ into (\ref{beta_n}) gives us the explicit bound
\begin{equation}
\label{beta_star}
\beta^\ast = \frac{2e(\sqrt{\logerror}+\sqrt{1+\logerror})^2 \ln 2}{(1-e^{-2})(1-2q)^2} \approx \frac{4.36(\sqrt{\logerror}+\sqrt{1+\logerror})^2}{(1-2q)^2}.
\end{equation}

\section{Simulation}
\label{sec:Simulation}



%
%

\subsection{Noisy Random Incidence Algorithm}

%

\begin{figure}
\begin{center}
\includegraphics[width=90mm]{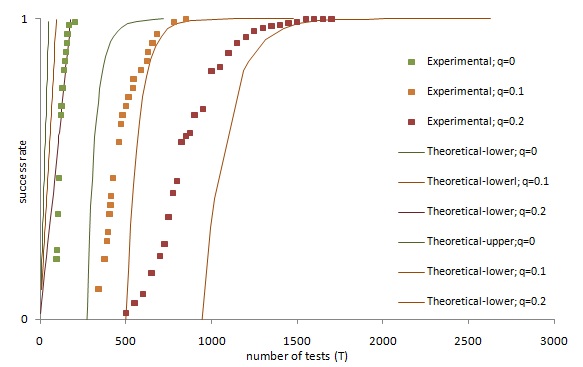}
\caption{The probability of success for Noisy-COMP as a function of the number of tests $T$, for different values of the noise parameter $q$.}
\label{plot3}
\end{center}
\end{figure}

We performed extensive simulations to validate our theoretical analysis. In the interest of space we present only Figure~\ref{plot3}, which examines the probability of error of Noisy-COMP as a function of the number of tests. Note that the experimental values we obtain correlate well with the corresponding bounds.


\begin{thebibliography}{10}
\providecommand{\url}[1]{#1}
\csname url@samestyle\endcsname
\providecommand{\newblock}{\relax}
\providecommand{\bibinfo}[2]{#2}
\providecommand{\BIBentrySTDinterwordspacing}{\spaceskip=0pt\relax}
\providecommand{\BIBentryALTinterwordstretchfactor}{4}
\providecommand{\BIBentryALTinterwordspacing}{\spaceskip=\fontdimen2\font plus
\BIBentryALTinterwordstretchfactor\fontdimen3\font minus
  \fontdimen4\font\relax}
\providecommand{\BIBforeignlanguage}[2]{{%
\expandafter\ifx\csname l@#1\endcsname\relax
\typeout{** WARNING: IEEEtran.bst: No hyphenation pattern has been}%
\typeout{** loaded for the language `#1'. Using the pattern for}%
\typeout{** the default language instead.}%
\else
\language=\csname l@#1\endcsname
\fi
#2}}
\providecommand{\BIBdecl}{\relax}
\BIBdecl

\bibitem{Dorfman1943}
R.~Dorfman, ``The detection of defective members of large populations,''
  \emph{Annals of Mathematical Statistics}, vol.~14, no. 436-411, 1943.

\bibitem{CGT}
D.-Z. Du and F.~K. Hwang, \emph{Combinatorial Group Testing and Its
  Applications}, 2nd~ed.\hskip 1em plus 0.5em minus 0.4em\relax World
  Scientific Publishing Company, 2000.

\bibitem{saligrama09}
G.~Atia and V.~Saligrama, ``Boolean compressed sensing and noisy group
  testing,'' \emph{CoRR}, vol. abs/0907.1061, 2009.

\bibitem{Macula1997217}
A.~J. Macula, ``Error-correcting nonadaptive group testing with de-disjunct
  matrices,'' \emph{Discrete Applied Mathematics}, vol.~80, no. 2-3, pp. 217 --
  222, 1997.

\bibitem{GV}
E.~N. Gilbert, ``A comparison of signaling alphabets,'' \emph{Bell System
  Technical Journal}, vol.~31, pp. 504--522, 1952.

\bibitem{Shannon:48}
C.~E. Shannon, ``A mathematical theory of communication,'' \emph{SIGMOBILE Mob.
  Comput. Commun. Rev.}, vol.~5, pp. 3--55, January 2001.

\bibitem{Sejdinovic2010}
D.~Sejdinovic and O.~Johnson, ``Note on noisy group testing: Asymptotic bounds
  and belief propagation reconstruction,'' in \emph{Communication, Control, and
  Computing (Allerton), 2010 48th Annual Allerton Conference on}, Oct. 2010,
  pp. 998 --1003.

\bibitem{NGT}
D.-Z. Du and F.~K. Hwang, \emph{Pooling designs and nonadaptive group testing:
  important tools for DNA sequencing}.\hskip 1em plus 0.5em minus 0.4em\relax
  World Scientific Publishing Company, 2006.

\bibitem{DyaRyk82}
A.~G. Dyachkov and V.~V. Rykov, ``Bounds on the length of disjunctive codes,''
  \emph{Probl. Peredachi Inf.}, vol.~18, pp. 7--13, 1982.

\bibitem{Dyachkov1989}
V.~V.~R. A.~G.~Dyachkov and A.~M. Rashad, ``Superimposed distance codes,''
  \emph{Problems Control Inform. Theory}, vol.~18, pp. 237 -- 250, 1989.

\bibitem{Cheng2009:1862-4472:297}
L.~G. Cheng~Yongxi, Du Ding-Zhu, ``On the upper bounds of the minimum number of
  rows of disjunct matrices,'' \emph{Optimization Letters}, vol.~3, 2009.

\bibitem{Chen20091581}
H.-B. Chen and H.-L. Fu, ``Nonadaptive algorithms for threshold group
  testing,'' \emph{Discrete Applied Mathematics}, vol. 157, no.~7, pp. 1581 --
  1585, 2009.

\bibitem{CandesTao}
E.~J. Cand\`{e}s, J.~K. Romberg, and T.~Tao, ``{Stable signal recovery from
  incomplete and inaccurate measurements},'' \emph{Communications on Pure and
  Applied Mathematics}, vol.~59, no.~8, pp. 1207--1223, Aug. 2006.

\bibitem{Donoho}
D.~Donoho, ``Compressed sensing,'' \emph{Information Theory, IEEE Transactions
  on}, vol.~52, no.~4, pp. 1289 --1306, April 2006.

\bibitem{Tropp07signalrecovery}
J.~A. Tropp, Anna, and C.~Gilbert, ``Signal recovery from random measurements
  via orthogonal matching pursuit,'' \emph{IEEE Trans. Inform. Theory},
  vol.~53, pp. 4655--4666, 2007.

\bibitem{springerlink:10.1007/BF01609876}
A.~Macula, ``Probabilistic nonadaptive group testing in the presence of errors
  and dna library screening,'' \emph{Annals of Combinatorics}, vol.~3, pp.
  61--69, 1999.

\bibitem{SOBEL01041975}
M.~Sobel and R.~M. Elashoff, ``Group testing with a new goal, estimation,''
  \emph{Biometrika}, vol.~62, no.~1, pp. 181--193, 1975.

\bibitem{CoverandThomas}
T.~Cover and J.~Thomas, \emph{Elements of Information Theory}.\hskip 1em plus
  0.5em minus 0.4em\relax John Wiley and Sons, 1991.

\bibitem{wiki:ccp}
W.~Feller, \emph{{An Introduction to Probability Theory and Its Applications,
  Vol. 1, 3rd Edition}}, 3rd~ed.\hskip 1em plus 0.5em minus 0.4em\relax Wiley,
  Jan. 1968.

\end{thebibliography}
\end{document}